\documentclass[article]{jdfsl}

\usepackage{makeidx}  
\usepackage{upgreek}

\usepackage{url}

\usepackage{amsmath}

\usepackage[T1]{fontenc}

\usepackage{array}

\usepackage{subfig}
\usepackage{multirow}
\usepackage{array}
\usepackage{amssymb}

\usepackage{natbib}

\usepackage{enumitem}

 \usepackage{amssymb}
 \usepackage{pifont}
 \newcommand{\cmark}{\ding{51}}%
 \newcommand{\xmark}{\ding{55}}%
\usepackage{textcomp}

\usepackage{upgreek}

\usepackage{url}

\usepackage{amsmath}
\usepackage{graphicx}

\usepackage[T1]{fontenc}

\usepackage{array}

\graphicspath{{./pdf/}{./jpeg/}{./images/}}
   \DeclareGraphicsExtensions{.pdf,.jpeg,.jpg,.png}

\begin{document}

\title{HTML5 Zero Configuration Covert Channels: Security Risks and Challenges}

\author{Jason Farina~~~~~~~~~~~~ \and Mark Scanlon~~~~~~~~~~~~ \and Stephen Kohlmann
 \and Nhien-An Le-Khac \and Tahar Kechadi}


\institute{~ \\ School of Computer Science \& Informatics,\\
University College Dublin, Ireland.\\
\{{jason.farina, stephen.kohlmann\}@ucdconnect.ie, \{mark.scanlon, an.lekhac, tahar.kechadi\}@ucd.ie}
}

\abstract{
In recent months, a significant number of secure, cloudless file transfer services have emerged. The aim of these services is to facilitate the secure transfer of files in a peer-to-peer (P2P) fashion over the Internet without the need for centralised authentication or storage. These services can take the form of client installed applications or entirely web browser based interfaces. Due to their P2P nature, there is generally no limit to the file sizes involved or to the volume of data transmitted -- these limitations will purely be reliant on the capacities of either end of the transfer. By default, many of these services provide seamless, point-to-point encryption to their users. The cyberforensic consequences of the potential criminal use of such services are significant. The ability to easily transfer encrypted data over the Internet opens up a range of opportunities for illegal use to cybercriminals requiring minimal technical know-how. This paper explores a number of these services and provides an analysis of the risks they pose to corporate and governmental security. A number of methods for the forensic investigation of such transfers are discussed.
}


\maketitle

\section{Introduction}

Sending anything larger than a small amount of data electronically is still a cumbersome task for many Internet users when reliant on popular online communication methods. Most email providers will limit the file size of attachments to something in the order of megabytes. Sending larger files usually requires users to upload the content to third party storage providers, e.g., Dropbox, OneDrive, box.net, etc., and provide a link to the content to their intended recipients. From a security standpoint, this leaves user vulnerable to their communication being intercepted or duplicated and their data being downloaded by others. Regarding the security of their data stored on this third-party provider, users must blindly trust this third-party to not access or share their data with any unauthorised party. 

While the requirement is ever increasing to send larger volumes of information over the Internet, the potential for third-party interception/recording of this data has become a common story in the general media. Recent leaks from whistle-blowers regarding the degree of surveillance conducted by large government funded spying agencies on everyday citizens has pushed the topic of cybersecurity into the public realm. Increasingly, Internet users are becoming conscious of their personal responsibility in the protection of their digital information. This results in many users being discontent with their personal data stored on these third party servers -- likely stored in another jurisdiction.

To respond to this demand a number of file exchange/transfer services have appeared in recent months facilitating the secure transfer of files in a peer-to-peer (P2P)fashion from point A to point B. Most of these services afford the user encrypted end-to-end file transfer and add an additional level of anonymity compared to regular file transfer services, e.g., email attachments, FTP or file sharing functionality built into most instant messaging clients. The more security conscious users will opt for the cloudless versions of these services. Opting for this level of control over personal information has upsides and downsides for the end user. The upside is that the user has precise knowledge over who has access to his/her information and what country the data is stored in. The downside comes in terms of reliability. The data stored or transferred using these services is only available if at least one host storing the file is online.

As with most security or privacy enhancing Internet services, these services are open to abuse by cybercriminals. In effect, the additional level of anonymity and security provided by these services provides cybercriminals with ``off-the-shelf'' counter-forensic capabilities for information exchange. Cybercriminal activities such as data exfiltration, the distribution of illicit images of children, piracy, industrial espionage, malicious software distribution, and can all be aided by the use of these services. 

\subsection{Contribution of this work}
\label{contribution}

For many, the topic of covert channels immediately brings to mind some form of steganography likely in combination with an Internet anonymising service, such as Tor and I2P. While some work has been conducted on the reverse engineering/evidence gathering of these anonymising P2P proxy services, little work has been done in the area of online services providing end users with the ability to securely and covertly transfer information from peer to peer in an largely undetectable manner. This work presented as part of this paper examines a number of popular client application and web based services, outlines their functionality, discusses the forensic consequences and proposes a number methods for potentially retrieving evidence from these services.

\section{Background Reading}
In order 

\subsection{Anonymising Services}
\label{anonymising}
Today there are many anonymising services available for communication and data transfer. The popular anonymous browser Tor allows users to explore the Internet without the risk of their location or identity becoming known \citep{loesing2010case}. The Tails operating system which works in conjunction with Tor offers an extra layer of anonymity over traditional operating systems. When a user is operating Tails all connections are forced to go through the Tor network and cryptographic tools are used to encrypt the users data. The operating system will leave no trace of any activity unless explicitly defined by the user.

The Invisible Internet Project, also known as I2P is another anonymous service similar to Tor. As I2P is designed as an anonymous network layer users can utilise their own applications on the network. Unlike Tor the I2P network traffic stays on the network. I2P does not use the traditional IP/Port user identification process but instead replaces this process with a location-independent identifier. This process decouples a user's online identity and physical location \citep{timpanaro2014group}. 

Both Tor and I2P provide anonymity to the user with an open network of onion routers. As the network of onion routers is run by volunteers it is continually growing . The result of this network growth is an increase in anonymity and privacy for each individual user \citep{herrmann2011privacy}.

\subsection{Untrusted Remote Backup}
\label{untrusted}
The MAIDSafe network is a P2P storage facility that allows members to engage in a data storage exchange. Each member of the network enables the use of a portion of their local hard drive by other members of the network. In return the member is given the use of an equivalent amount of storage distributed across the network and replicated to multiple locations referred to as Vaults. This allows the authorised member access from any location and resilience should a portion of the network not be active at any time. All data stored on the network is deduplicated and replicated in real time with file signatures to ensure integrity. In addition the data is encrypted allowing secure storage on untrusted remote systems. Authorised access is managed through a two factor authentication (password and pin). The use of the MAIDSafe network is incentivised through SafeCoin, a cryptocurrency that members can earn by renting out space or providing resources such as bandwidth for file transfers. Other users can earn SafeCoins by participating in development of the protocol.

\subsection{Types of File Transfer Attacks}
\label{risk}

Data exfiltration refers to the unauthorised access to otherwise confidential, proprietary or sensitive information. \citet{giani2006data} outlines a number of data exfiltration methods including most regular file transfer methods for ``inside man'' attacks, e.g, HTTP, FTP, SSH and email, and external attacks including social engineering, botnets, privilege escalation and rootkit facilitated access. Detection of most of these methods is possible using a combination of firewalls and network intrusion detection systems or deep packet inspection \citep{liu2009sidd, sohn2003study, cabuk2009ip}.


\subsection{The Deep-Web}
The Deep-web refers to layers of internet services and communication that is not readily accessible to most users and is not crawled by search engine crawlers. Unlike the Internet there is no one set of protocols or formats for the Deep-Web, instead deep-web is a generic term used to describe internet communications that are managed using closed or somehow restricted protocols. More recently however, the term ``deep web'' has been made synonymous with black-market sites such as ``The Silk Road'', taken offline by the FBI in 2013.

\subsubsection{The FreeNet Project}
The FreeNet project is a peer based distributed Internet alternative. Users connect to FreeNet through an installed application that uses multiple encrypted connections to mask the source and destination IP addresses as well as the traffic and data location itself. FreeNet peers, or nodes, store fragments of data in a distributed fashion. The number of times a data item is replicated is dependant on the demand for that data. More popular files have more available sources resulting in better availability and faster access times for the requesters. The layered encryption of the connections provides an effective defence against network sniffing attacks and also complicates network forensic analysis. FreeNet by default was deployed in OpenNet configuration, meaning connections could be made with a random set of nodes from all of the FreeNet nodes available. Users can choose to instead use a DarkNet configuration where only known nodes are routed through resulting in a small world topology for the network. The choice of OpenNet or Darknet is not binary however as the user can choose to employ a mix of both to whatever degree they wish however small world topology with random connections may result in inefficiencies. DarkNet routing tables are created using location swapping to populate efficient routes based on relative node distance . This feature gave rise to an attack, Pitch Black, which caused degradation of DarkNet performance and security by tricking the network into moving disproportionate amounts of data to individual peers by falsely reporting node distances to corrupt the routes \citep{evans2007routing}.\\
The Pitch Black attack will only work against a DarkNet configuration but OpenNet is inherently less secure than DarkNet. In OpenNet, any user can, theoretically, enumerate and connect to all available nodes and perform traffic correlation to trace back to the original requester. This ability to harvest addresses and to identify the initial seed nodes results in OpenNet configurations being very easy to block with a firewall by blacklisting a dynamic list of known nodes and DarkNet is the recommended setup for users to employ.

\subsubsection{Tor}
The Tor Network is another networking protocol designed to provide anonymity. Initially Tor, an acronym for ``The Onion Router'', just provided random routes with three encrypted connections before exiting through an exit node. Exit nodes are selected based on the resources available at the exit node and the destination port of the users traffic. If the exit node is not configured to accept traffic destined for a service's port it will not accept any routing requests bound for that port. 

Tor works by creating generating a circuit one relay at a time between Tor nodes before emerging to the internet through the exit nodes mentioned above. Each relay only knows the system it is receiving data from and the server that it is sending data on to next. Once the circuit is established all traffic for that particular session is routed along the same path. Should the user decide to use a different service or visit a different website, a new circuit would be generated and used. Tor is a SOCKS (Socket Secure) proxy meaning any TCP application that can be made SOCKS aware can be used through the Tor network. Software that is not SOCKS aware, like the Gnutella P2P protocol for example, will run through Tor but will not properly anonymise their traffic. A Gnutella peer will still send its true IP address as part of any QueryHit message. The same is true for any BitTorrent Sync (a popular cloudless Dropbox alternative) traffic attempted through Tor.

Sometimes a user can opt to not utilise an exit node and instead browse content on services hosted within the Tor network itself. These web pages, such as the Silk Road, are known as Hidden Services and are anonymised systems just like the clients visiting. This anonymisation would normally make any form of dependable routing impossible as any DNS lookup would not be able to report a static IP address for the service for the client system to connect to. Instead, hidden services use a distributed hash table (DHT) and a known service advertising node to inform clients of their presence and the path to the service provided. The process of making a Hidden Service available is as follows:

\begin{enumerate}
\item{The Hidden Service (HS) picks a public ``identity key'', \texttt{$ I_s$}, and an associated security key \texttt{$S_s$}}.
\item{An Onion Identifier \texttt{$O_s$} is generated as the function $H(I_s)$ where H is the first 80 bits of the 160-bit SHA1 hash base32 encoded. This gives a 16 byte string that then has the TLD \texttt{.onion} attached}
\item{ HS constructs three circuits each with 3 relays but no exit node. These relays will act as introduction nodes to the service.}
\item {A descriptor is then signed by the HS with the secret key \texttt{$S_s$} and the Public \texttt{$I_s$} along with the relay IPs are published to the DHT using a time period as an index.}
\item{The visiting client system requests the descriptor from the DHT and generates a circuit to one of the Introduction relays. }
\item{ Once the connection is established the Client and the Introduction system agree on a rendezvous relay \texttt{RP} for both the server and the client to use as a common relay to create a bridge}
\item{Introduction relay informs the HS of the RP address and both the HS and the client set up circuits meeting at the RP}
\end{enumerate}

Due to the HS reliance on DHT and HS Directory servers to perform the introductions for Tor users, \citet{Biryukov2013} were able to demonstrate a denial of service attack on a HS by impersonating the HS directory servers acting as Introduction nodes. They were also able to crawl the DHT to harvest Onion Identifiers over a period of 2 days to provide a more accurate index of the content of the DeepWeb contained within Tor. This approach could be useful when attempting to stall a HS until it can be properly identified. Botnet C\&C nodes could be prevented from issuing updates or harvesting reports from the bots through the methods outlined in their paper.

\subsection{File Sharing Services Built on Anonymising Networks}
\label{onions}
OnionShare is a file sharing application that leverages the anonymity of Tor to provide secure file transfers for its users. File transfers are direct from uploader to recipient though both users utilise the Tor browser to participate. OnionShare itself is a python based application that sets up a fileshare on the local system as a limited web server. This web server is then advertised as a Tor Hidden Service using the built in functionality of the Tor browser. The application uses random data to generate a 16 character onion address and more random data to generate a unique name for the file being shared to use as a reference for.

The process used by OnionShare is as follows:

\begin{enumerate}
\item Uploader starts Tor Browser to provide an entry point for OnionShare to the Tor network.

OnionShare is started and a temporary directory is created in the users' default temp folder. All randomly generated names in OnionShare follow the same procedure:
\begin{enumerate}
\item{A number of random bytes are generated using os.random. 8 are generated for a directory/host name and 16 for the filename "slug" used to generate the file portion of the share URL}
\item{ These random bytes are SHA-256 and the rightmost 16 characters of the resulting hash are carved }
\item{h is then base32 encoded, all characters are converted to lower case and any trailing `=' signs are removed}
\end{enumerate}

\item The result is then used as a URL using the format \texttt{<host>.onion/<fileID>} and this is the url the Tor browser advertises to the introduction nodes and registers on DHT.

\item The uploader then sends the URL to the downloader who must use the URL within a timeframe (24 hours by default) or the signature of the file HS timestamp will not match, a process controlled by the ItsDangerous library for python. In addition to this time limit, OnionShare also utilises a download counter which has a default value of 1. Once the number of downloads successfully initiated matches this counter the link is no longer considered valid and all incoming URLs with the sae file signature are refused.

\end{enumerate}

This combination of time and availability in conjunction with the anonymity of Tor HS makes OnionShare traffic extremely difficult to analyse effectively. If the traffic is observed then the link is already invalidated. Similarly, if the file is discovered on a local filesystem by an investigator.
 


\section{Investigative Techniques}
\label{related}

While no directly related work has been published at the time of writing, there are a number of digital evidence acquisition methods published for related services. This section outlines a number of related investigation techniques and analyses their relevancy to the forensic recovery of evidence from covert file transfer services.

%

\subsection{Cloud Storage Forensics}
\label{cloudsotrageforensics}

Forensics of cloud storage utilities can prove challenging, as presented by \citet{Chung201281}. The difficulty arises because, unless complete local synchronisation has been performed, the data can be stored across various distributed locations. For example, it may only reside in temporary local files, volatile storage (such as the system's RAM) or dispersed across multiple datacentres of the service provider's cloud storage facility. Any digital forensic examination of these systems must pay particular attention to the method of access, usually the Internet browser connecting to the service provider's storage access page (https://www.dropbox.com/login for Dropbox for example). This temporary access serves to highlight the importance of live forensic techniques when investigating a suspect machine as a ``pull out the plug'' anti-forensic technique would not only lose access to any currently opened documents but may also lose any currently stored sessions or other authentication tokens that are stored in RAM. 


\cite{Martini2013287} published the results of a cloud storage forensics investigation on the ownCloud service from both the perspective of the client and the server elements of the service. They found that artefacts were found on both the client machine and on the server facilitating the identification of files stored by different users. The module client application was found to store authentication and file metadata relating to files stored on the device itself and on files only stored on the server. Using the client artefacts, the authors were able to decrypt the associated files stored on the server instance.

\subsection{Network Forensics}
\label{network}

Network Forensic Analysis Tools (NFATs) are designed to work alongside traditional network security practises, i.e., intrusion detection systems (IDSs) and firewalls. They preserve a long term record of network traffic and facilitates quick analysis of any identified issues \citep{corey2002network}. Most firewalls allow HTTP and HTTPS traffic through to allow users behind the firewall to have access to regular web services which operate over these protocols. With regards to the web based covert file transfer services (outlined in detail in Section\ref{webbased} below), blocking network traffic to these systems would require the maintenance of a comprehensive firewall list of such servers to ensure no unauthorised data exfiltration. NFATs collecting this information will only have the ability to capture the encrypted packets, their destination and associated metadata. Identifying precisely what has been transferred will likely prove impossible for network administrators.

The issue with always-on active network forensics is dealing real-time with the large volumes of traffic involved. One approach to overcome the massive volume of network data to process is to simply record every packet sent and received from the Internet, in the similar manner to the tactic employed in the case outlined by \citet{garfinkel2002network}. This would facilitate an after-the-fact reconstruction of any data breaches to aid in determining precisely what was compromised.

\subsection{Deep-Web Forensics}
\label{deepweb}
Although Tor and I2P are designed for users to communicate and transfer data on the Internet anonymously it is still viable and possible for investigators to gather user specific information. The process of an investigation into the Tor network requires advanced digital forensic knowledge as traditional investigation methods used for standard networks fail to heed the desired results. \citet{loesing2010case} published a study measuring statistical data in the Tor network. The study is weighted towards protecting the users anonymity while using Tor but nonetheless shows that it is technically possible to gather data on Tor users by setting up a Tor relay and logging all relayed user traffic. 

When users install Tor the software first connects to one of the directory authorities. The directory authorities are operated by trusted individuals of Tor and from these authorities the Tor software downloads the list of currently available Tor nodes. These nodes are relay servers that are run by volunteers of Tor. The Tor client then selects three nodes from those available and builds an encrypted channel to the entry node. An encrypted channel is then built from the entry node to the middle node, and lastly this channel connects to the exit node. 

The exit node is not aware of the entry node or who the client is and the entry node does not know which exit node is ultimately selected by the client. It is possible to de-anonymise streams by only recording every 1 in every 2000 packets \citep{murdoch2007sampled}.

Although Loesing et al. demonstrate the process of exposing traffic by port and countries of connecting clients the results remain masked. The aim of the work was to outline a process by which researchers can safely measure network data in anonymity systems while protecting the privacy of the user.

\citet{blond2011one} demonstrated the results of an attack on the Tor anonymity network that revealed 10,000 IP addresses over 23 days. The authors used the attack to obtain the IP addresses of BitTorrent users on Tor. The study found that 72 percent of users were specifically using Tor to connect to the tracker. The authors launched their attacks through six instrumented Tor exit nodes resulting in 9 percent of all Tor streams being traced. Moreover, the paper analyses the type of content discovered in the attack culminating in the result that the existence of an underground BitTorrent ecosystem existing on Tor is plausible. Alongside these attacks Tor users were also profiled. Using BitTorrent as the insecure application they hijacked the statistical properties of the DHT and the tracker responses.


\section{Evolution of HTML5 Powered Covert File Transfer Services}
\label{webbased}

Preamble

\subsection{Basic HTML5 File Transfer}
\begin{figure}[!t]
\centering
\includegraphics[width=0.3\textwidth]{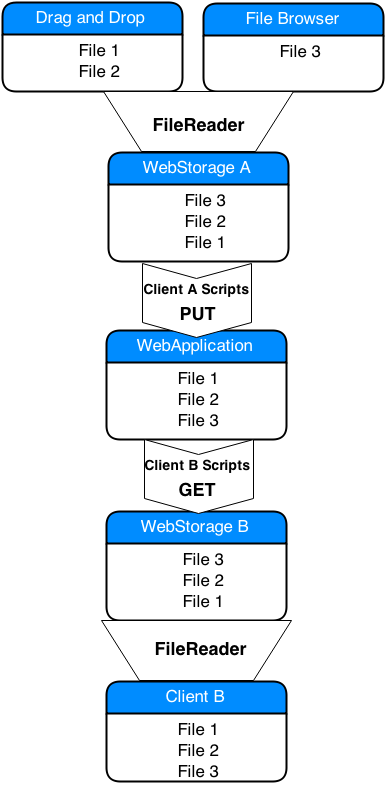}
\caption{The basic HTML5 data transfer process}
\label{fig:dataflow}
\end{figure}

Basic HTML5 File transfer as depicted in figure\ref{fig:dataflow} is accomplished using native browser APIs that allow a user to utilise a \texttt{data transfer} object. This object consists of a customisable array of key:value pairs that represent a group of file objects. This associative array is then accessible by client side scripts run from a web page or web application. These scripts must first be downloaded and allowed to run by the local user (this depends on the trust setting for the website being visited). Any element can be added to the array through a Drag and Drop (DnD) functionality or files can be added though a file browser interface. The actions available by default are:
\begin{itemize}
\item copy: A copy of the source item may be made at the new location.
\item move: An item may be moved to a new location.
\item link: A link may be established to the source at the new location.
\item copyLink: A copy or link operation is permitted.
\item copyMove: A copy or move operation is permitted.
\item linkMove: A link or move operation is permitted.
\item all: All operations are permitted.
\item none: The item may not be dropped
\end{itemize}

if the element added to the array is a file then the element is passed to a \texttt{FileReader} object that copies the data contained in the file to \texttt{localstorage} or \texttt{session storage} depending on the settings of the web application. Local Storage is shared across all browser sessions currently active, session storage is only available to the owning application or window (for browsers with multiple tabs or windows). This local/session storage behaves very similarly to the standard cookie storage but with hugely increased capacity. ( 5MB for Chrome, Firefox and Opera, 10MB for Internet Explorer - DnD native is only available in version 9+ of IE - web storage in version 8+, and 25MB for Blackberry).

For basic Data transfer, the \texttt{Filereader} reads the entire file indicated into RAM for processing. Once stored in webstorage a file can only be accessed by local actions that have permission to access that web storage area such as client side scripts downloaded from the controlling web page or session. These scripts can use any scripting language but usually JQuery, Javascript or AJAX. The local client can also call scripts to run on the remote server in order to pass variables or prepare for the establishment of additional sessions are required.

\subsection{Cryptographically enhanced HTML5 data channels}
Following on from their acquisition of ON2 in February 2010, Google continued to develop a browser to browser data transfer protocol that was made open source in 2011 when it was adopted by W3C as a standard for HTML5. The protocol, which supported Real Time Communication between browsers was released as WebRTC 1.0, was developed to provide P2P voice, video and data transfers between browsers without an additional software requirements. WebRTC, provides a collection of protocols and methods as well as a group of codec libraries that can be accessed via a javascript API . The specification for WebRTC can be found on the W3C developer site \url{http://www.w3.org/TR/webrtc/}. 

WebRTC improved data transfer over the standard HTML5 script based method by introducing data integrity, source authentication and end to end encryption. This is accomplished through the use of Datagram Transport Layer Security (DTLS) \cite {modadugu2004design} extension to handle key exchange for the Secure Real-time Transport Protocol (SRTP). The use of DTLS-SRTP differs from standard VOIP encryption by removing the need to trust SIP relays that form the path between the source and destination.

\begin{figure}[!t]
\centering
\includegraphics[width=0.3\textwidth]{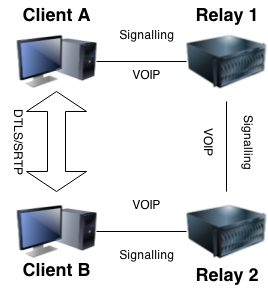}
\caption{traditional VOIP data vs DTLS-SRTP}
\label{fig:dtls}
\end{figure}

In the image \ref{fig:dtls} the standard VOIP method of communication is displayed alongside the newer WebRTC method. Both systems start with establishing a signalling and control path to handle non-sensitive data such as connection auditing packets. In VOIP the data stream would follow this established path and traffic between Client A and Client B involving relay through Relays 1 and 2. Unless the user fully trusts both relays and the security of the network path between each node on the network, there is a risk of an adversary eavesdropping on the data stream and either manipulating the content in transit or capturing it for offline inspection.

Under WebRTC, the signalling path is still used but through the use of Interactive Connectivity Establishment (ICE) the protocol initiates a direct connection between the endpoint clients allowing data to be transferred directly while the control signalling continues to follow the original path. To allow for clients obfuscated by firewalls that perform NAT operations to present a public facing IP WebRTC uses Session Traversal Utilities for NAT (STUN) to allow NAT traversal so private IP ranges can connect directly to to one another across public networks. Should STUN fail due to an inability for direct connection, WebRTC will employ Traversal Using Relay NAT (TURN protocol) which will designate one trusted intermediary server to act as a relay for the Client traffic.

\section{Analysis of Existing Services}
\begin{table*}[!t]

\centering
\begin{tabular}{|l|c|c|c|c|c|c|c|c|}
\hline
\textbf{Service Name}                     &\rotatebox {75} {\textbf{Encrypted Transfer}}   & \rotatebox {75} {\textbf{HTML Based}}   & \rotatebox {75} {\textbf{Registration Required}} & \rotatebox {75} {\textbf{Application Option}} & \rotatebox {75} {\textbf{Anonymity}}    & \rotatebox {75} {\textbf{Mobile Compatibility}} & \rotatebox {75} {\textbf{Persistance Storage}} & \rotatebox {75} {\textbf{Relay Server}}        \\ \hline
Sharefest    & \cmark & \cmark & \xmark  & \xmark        & \xmark & \xmark          & \cmark        & \cmark \\ \hline
PipeBytes          & \cmark & \cmark & \xmark  & \cmark        & \cmark & \xmark          & \xmark        & \cmark \\ \hline
JustBeamIt       & \xmark & \cmark & \xmark  & \xmark        & \xmark & \cmark          & \xmark        & \xmark \\ \hline
Transfer Big Files & \cmark & \cmark & \cmark  & \xmark        & \xmark & \cmark          & \cmark        & \cmark \\ \hline
Infinit          & \cmark & \cmark & \cmark  & \cmark        & \xmark & \xmark          & \cmark        & \cmark \\ \hline
Any Send       & \cmark & \cmark & \cmark  & \cmark        & \xmark & \cmark          & \cmark        & \xmark \\ \hline
Rejetto        & \xmark & \xmark & \cmark  & \cmark        & \xmark & \xmark          & \cmark        & \cmark \\ \hline
QikShare         & \xmark & \cmark & \cmark  & \xmark        & \xmark & \cmark          & \cmark        & \cmark \\ \hline
AeroFS           & \cmark & \cmark & \cmark  & \cmark        & \cmark & \cmark          & \cmark        & \cmark \\ \hline
\end{tabular}
\caption{Comparison of Browser Based Transfer Services\label{tabcomparison}}
\end{table*}

Free services such as PipeBytes and Any Send \url{http://getanysend.com} as outlined in Table~\ref{tabcomparison} allow a user to transfer files of any size via the browser without the need for external software. There are many similar services that a user can access easily online and this abundance of services is of particular interest to digital forensic investigators. As can be seen in Table~\ref{tabcomparison}, PipeBytes and AeroFS also offer anonymity to there users. AeroFS also offer a private cloud whereby absolutely no data or communication is transferred or stored on the AeroFS servers. Although AeroFS is primarily marketed as a business service it is possible for this private cloud to be used for illicit and illegal transfers. Services such as QikShare and Transfer Big Files have added the challenge of deciphering mobile compatibility for the digital forensic investigator. Multiple device and cross platform file sharing offers great speed and flexibility to the standard user. The downside to this is not the security of these file transfers but it is the lack of a tool that can sift out the illegal and illicit data transfers. All services listed in Table~\ref{tabcomparison} with the exception of PipeBytes and JustBeamIt offer persistence storage. Although some of the services only offer this as a paid option it is quite feasible for a user to select these services to store incriminating material without risking a trace on the users local machine. Table~\ref{tabcomparison} is not extensive and it is important to note that there are a multitude of these services available online.

\subsection{HTML5 Enabled Peer-to-Peer Transfer}

sharefest.me is a file-sharing ``one-to-many'' based website that aims to dynamically generate and maintain file-sharing swarms by connecting peers that are interested in sharing the same data. Like the BitTorrent protocol, multiple peers are utilised simultaneously to transfer portions of the data thus increasing download speeds by avoiding the bottleneck that is the lower upload speed of a standard ADSL internet connection. In order to achieve this, Sharefest is built on Peer5's (\url{https://peer5.com/}) platform for a distributed internet, a P2P data transfer mesh network that utilises the capabilities of the browser without additional plugins beyond a WebRTC capable browser. WebRTC is a HTML5 integrated utility for providing data channels for real time communication such as VOIP or IM but which can also be used to transfer static data files. WebRTC is handled by default in Chrome (version 33+), Mozilla (version 31+) and Opera (version 25+) desktop browsers as well as Android browser (version 37) and Chrome for Android (versions 38+). 

\begin{figure}[!t]
\centering
\includegraphics[width=0.3\textwidth]{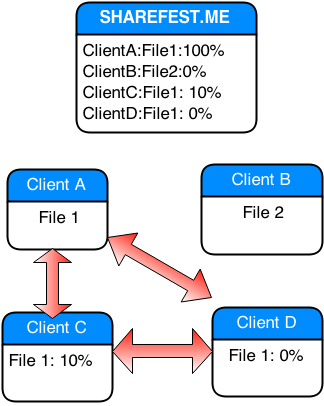}
\caption{Sharefest P2P mesh over WebRTC}
\label{fig:mesh}
\end{figure}

As depicted in figure \ref{fig:mesh} the Sharefest process is quite straightforward in design. The sharefest.me server acts as a transfer control server that records all files being offered for sharing and matches the resource to the client system request. In the scenario depicted, Client A has a complete file that it wants to share. Client A connects to the Sharefest server at \url{https://www.sharefest.me/} over port 443 and negotiate TLS1.2 where possible using SPDY if available for web content transfer. Given a full range of options the Sharefest server negotiates the use of the ECDHE-ECDSA with AES 128 and GCM 256. As required by the IETF RFC 4492 (\url{http://tools.ietf.org/html/rfc449}) the Sharefest server passes the Curve details as part of its serverkeyexchange packet.

Once a secure path is established the server delivers a small set of helper scripts to the client:
\begin{itemize}
\item files.js : a script to gather file details from the client
\item ui.js : a script to control the update and display of the file management interface on the page.
\end{itemize}

Once a file has been selected for sharing the Sharefest server assigns a code value in the form of a URL such as \url{https://www.sharefest.me/67509cb244257b6643540dda512f8171} where the number after the domain name is the swarmID for this file. The SwarmID has 32 characters but is not based on the MD5 of the file, instead it appears to be derived from the Sharefest crypto.js script which incorporates SHA3 in a lengthy calculation. The swarmID is deterministic meaning that any client sharing the same file will identified with the same swarmID. 

Once clients offering and requesting the same file or fileset are identified the Sharefest server acts as a central traffic control and initiates a STUN connection directly between the participating clients. In figure \ref{fig:mesh} Clients A C and D are all participating in the swarm for File 1 but Client B is not involved. The STUN connection consists of each pair of clients issuing BIND requests from uploader to downloader using STUN over UDP which allows each host ot discover its own public facing IP address in order to create end to end connections through a NAT gateway or firewall. Each BIND / Confirm pair is re-issued on a regular, short, interval to ensure the connection remains intact. Once the STUN session is active the two peers negotiate a WebRTC session and switch over to the protocol's encryption. ACK and STUN confirmation messages continue to be sent to and from the Sharefest server and the peers throughout the exchange.

Sharefest is an example of P2P privacy in a distributed network ensuring that data can be transferred without risk of interception. This level of privacy comes at a cost though as the ability of IT security to inspect the traffic is greatly diminished with the level of encryption in use at all stages of the transfer. Packet analysis can detect the IP addresses in use but without access to the key to decrypt the traffic the content being transferred is extremely difficult to determine. One option available to network admins is to block the use of the sharefest.me service by blacklisting the URL. This would have the effect of preventing casual usage of the service but the source for Sharefest is publicly available on Github \url{https://github.com/Peer5/Sharefest} along with instruction and support for installation of a personal server. Peer5 also provide the API key for free to anyone interested in the code. This means that any IP or URL could become a Sharefest server.

One method of detecting the use of this application is the STUN traffic generated once a peer is identified and connection is initiated. In testing an average of 5 STUN negotiation/confirmation exchanges were recorded every second depending on the level of file transfer data passing between the peers. This level of ``noise'' would make the user of Sharefest relatively easy to discover and no effort is made to obfuscate the communicating peers.

In an attempt to determine if this lack of anonymisation could be overcome, we attempted to run Sharefest through a Tor circuit but all three transfer utilities (JustBeamIt, PipeBytes and Sharefest) failed to complete the initial setup with the relevant server. This was tested using Tor Browser installed on a Windows 7 VMWare image. A possible alternative may be to attempt the use of a SOCKS aware proxy to direct the traffic to and from the application. Alternatively a server running Sharefest could be adapted to run as a Tor Service but this would not alleviate the lack of privacy experienced once data transfer was initiated between peers.

\subsection{JustBeamIt}

An example of a basic HTML 5 transfer application is the file transfer service offered at \url{http://www.justbeamit.com}. The sending user connects to the server over port \texttt{8080} (alternate HTTP port) and performs a standard TCP handshake followed by a series of \texttt{HTTP GET} requests for client side JavaScripts.
\begin{itemize}
\item JustBeamIt.js - the base script that sets up the variables and defines the communication functions
\item BrowserDetectUtility.js - determines if the user browser can properly support HTML5 data transfers
\item FileHandler.js - manages the transfer to and from the file array. Handles the array reset and webpage notifications if the array is emptied.
\item UploadManager.js - defines the drag and drop actions and defines the ``landing zone''
\item UploadHandler.js - Determines if the Client needs to use XMLHttpRequest (XHR) or FORM based uploading and generates the QRCode.
\item UploadHandler.XHR.js and UploadHandler.FORM.js - the actual uploading scripts
\end{itemize}

There is an option to drag and drop a file into the browser but in this instance the file browser is used to select a file from the local user pictures folder. Once selected the button ``Create Link'' is clicked and the link \url{http://www.justbeamit.com/di33x} is created along with a QRCode for mobile use. This link can copied and sent to the receiving system, in the meantime the local client is redirected to a relay server URL for the upload itself (\url{http://b1.justbeamit.com/}. On the remote system, the URL is pasted into a browser and the system and the remote client loads \url{http://www.justbeamit.com} and immediately requests the download token for the file di33x and downloads and runs the set of JavaScripts. The server passes along the token along with the current download status (upload waiting) and the file descriptor (name, size, extension). Once the remote user clicks on the link to download the browser is redirected to \url{http://b1.justbeamit.com} where the file transfer is performed. Once complete the local user is notified that the transfer has been successful. The token used to download is now invalidated and a new link must be generated if the file is to be shared again. Similarly, there is a 1000 second timeout period during which the shared file must be downloaded before the opportunity expires and a new share token must be generated.

While this method of file transfer provides ease of use to the end user, all transfers are performed via unencrypted traffic. The data being transferred is susceptible to any form of eavesdropping capable of detecting traffic on any network segment the traffic passes through. The open nature of the transfer and the client side execution of scripts (as well as the open exchange of tokens) allow for trivial man-in-the-middle (MITM) attacks where an adversary capable of eavesdropping can use a proxy or other interception utility to alter the packets in transit. One possible scenario would be the substitution of a harmless UploadManager.js script for something less benign as identified by Jang-Jaccard\cite{JangJaccard2014973} as a rising risk or, even exchanging the generated download token for one that leads to a virus or other form of malware.

From a security standpoint, defence against the use of this service is quite straightforward. Because of the application's use of a centralised set of servers, a standard firewall rule to block access to \url{http://*.justbeamit.com} would prevent any upload but also any attempt to download.

\subsection{PipeBytes}

A second system of this type is PipeBytes.com. Just as with JustBeamIt PipeBytes allows a user to drag and drop files into the browser or use a file browser to select files for upload. Also as with JustBeamIt, there is a standard connection to the PipeBytes.com server though this time it is over the standard \texttt{HTTP} port 80. Once connected the uploading client downloads a series of scripts to be run locally and the site itself is built on the freely available open source YUI API \url{http://yuilibrary.com/} . However, one immediate difference is that, as part of the packet traffic (which again is unencrypted) the client system uploads a random number generated using the \texttt{MATH.RANDOM} javascript function. This random number (between 0 and 1) is used as a parameter for a serverside script called from the upload-callbacks.js line:\\ YAHOO.util.Connect.asyncRequest('GET', 'getkey.php?r=' + Math.random(), callback) \\ the getkey.php page that is output from the server side script contains a key value that is used to populate the URL / Code box option on the webpage. The value of the key is use din subsequent scripts and is derived from o.responsetext. Once a key have been generated and the upload button is clicked, the local client issues a POST request for put.py on the server using the key and a second random number as parameters. The file is then uploaded to the server as an unencrypted transfer. While the system waits for the remote client to connect, the local uploader issues frequent GET requests for status updates.

Once the remote client receives the URL or code to initiate the download they connect to the same url (\url{http://www.PipeBytes.net} over port 80 to input their download code (the Key) where they will be redirected to one of the hosting servers (in testing only host03.PipeBytes.net was discovered though the naming convention would suggest that others do exist or have existed). The connection request takes the form of:\\ \texttt{$GET /get.php?key=103223658539867 HTTP/1.1$}\\ which is acknowledged with a standard HTTP 200 OK message from the server. The transfer of the file then commences and the traffic is delivered unencrypted.

While initially this version of the system appeared to provide more privacy and security by having the key generated and managed by scripts run on the server or the systems hosted by developer.Yahoo.com, the plain text transfers from server to peer leave it vulnerable to the same MITM attacks as justbeamit.com. In fact, any file upload system that uses open channels to transfer authentication data or transfers the data itself in the clear cannot be said to be either secure or private systems. For sensitive data transfers a more robust alternative was required.

\section{Forensic Consequences of ``Untraceable'' File Transfer}
\label{why}

The facilitation of ``untraceable'' or ``anonymous'' file exchange can lead to a number of potential malicious use cases. For each of the scenarios outlined below, an added dimension can be created by the originator of the content: time. Due to the ability to create ``one-time'' or temporary access to any piece of content, the timeframe where evidence may be recovered from remote sharing peers might be very short.

\subsection{Cybercriminal Community Backup} Unmonitored covert transfer could be used to create a ``share and share alike'' model for the remote encrypted backup of illegal content. The sharing of these backups onto multiple remote machines effectively could provide the user with a cloudless backup solution requiring minimal trust with any remote users. The encryption of the data before distribution to the community can ensure that only the owner will ever have access to decrypt the data. Trust only comes into play should the remote nodes delete the information or it otherwise becoming unrecoverable. Having a secure, encrypted connection to a remote backup might be desirable to cybercriminals enabling the use of a kill-switch to their local storage devices should the need arise. 

\subsection{Secure Covert Messaging} For example, the proof of concept based on the BitTorrent Sync Protocol found at \texttt{http://missiv.es/}. The application currently operates by saving messages to an ``outbox'' folder in a synchronised share between peers that has a read only key shared to the person you want to receive the message. They in turn send you a read only key to their outbox. One to many can be achieved by sharing the read only key with more than one person but no testing has been done with synchronisation timing issues yet and key management may become an issue as a new outbox would be needed for each private conversation required.

\subsection{Industrial Espionage} Many companies are aware of the dangers of allowing unmonitored traffic on their networks. However, quite often corporate IT departments enforce a blocking of P2P technologies through protocol blocking rules on their perimeter firewalls. This has the effect of cutting off any file-sharing clients installed on the LAN from the outside world. In addition to Deep Packet Inspection (DPI) to investigate the data portion of a network packet passing the inspection point, basic blocking of known IP address blacklists in firewall rulesets can be used. The difficulty in blocking HTTP based file transfers is that the technology is likely used during regular employee Internet usage. HTTP transfers can be used when emailing file attachments or adding items to content management system. One additional scenario where these services could be used would be to transfer files within a LAN and subsequent external exfiltration from a weaker/less monitored part of the network, e.g., guest wireless access.

\subsection{Piracy} Like any other P2P technology, the ability to transfer files in a direct manner from peer to peer lends itself well to the unauthorised distribution of copyrighted material. The sharing of copyrighted multimedia, software, etc., between peers using these covert services is less likely to lead to prosecution compared with public piracy on open file-sharing networks such as BitTorrent.

\subsection{Alternative to Server Based Website Hosting} This scenario involves the creation of static websites served through a shared archive. These websites could be directly viewed on each user's local machine facilitating the easy distribution of any illegal material. The local copies of the website could receive updates from the ``webmaster'' through the extraction of archived updated distributed in a similar manner as the original.

\subsection{Malicious Software Distribution} Due to the lack of any cloud based virus scanning, e.g., provided by most email providers and a number of file synchronisation providers, the direct unscanned transfer of files could facilitate the distribution of malware to regular Internet users.

\section{Potential Forensic Investigation Techniques}
\label{potential}

Assuming access (physical or remote) can be acquired to either end of the file transfer, then a live acquisition of the evidence should be attainable. Performing evidence acquisition after the fact would rely on traditional hard drive and memory forensic techniques to see if any remnants of the network communication remain.

The investigation of the unauthorised transfer for information through one of these services without access to either end of the transfer can prove extremely difficult. Assuming through some external means, the precise date and time of the transfer were discovered. The only method available to law enforcement is to effectively wiretap the transfer by running a software or hardware based deep packet inspection tool on the network at either end of the transfer.

To date there has been keen interest in research performed on the forensic examination of file sharing utilities and the type of security risks they pose. \citet{chung2012digital} outlined a best practise approach to the investigation of file sharing using cloud based storage-as-a-service (StaaS) utilities such as Dropbox, iCloud and One Drive. In 2014, \citet{Federici201430} presented Cloud Data Imager (CDI) a utility developed by the Italian Polizia to automate the retrieval of cloud based storage artefacts from a suspect system and use these credentials to access their secure storage online.

\citet{scanlon2014leveraging} described a methodology that leveraged the processes used by persistent file synchronisation services to ensure data integrity to retrieve data that would otherwise have been inaccessible. This could be as a result of deliberate obfuscation such as encryption or anti-forensic activities or it could be caused by an error in the imaging process. The methodology presented utilised the need for synchronisation group ongoing communication to enumerate remote peers and to identify any authorised peers that could provide a forensically true copy of the suspect data.

Emerging file transfer utilities, such as the purely browser based file transfer utilities based on WebRTC, do not advertise persistence of availability nor integrity checking beyond the initial transfer and in many cases are only associated for the length of time that both parties are online and in communication, directly or otherwise. After this time, such as with Onionshare for example, the address of the file source will change completely and no longer be available to any peer authorised or otherwise.

This ephemeral nature of data transfer can make any attempt to verify or re-create the circumstances of the file transfer difficult if not impossible and it very much depends on the features of the individual application being employed.

\section{Conclusion}
\label{conclusion}

The evolution of online file transfer systems is becoming more and more covert through employing encryption-based, server less, P2P protocols. The development of HTML5 and JavaScript based services proves particularly interesting from a digital forensic perspective. As these technologies mature, a future can be easily envisioned whereby the investigation and evidence retrieval from these systems will prove extremely difficult, if not entirely impossible. Daisy-chaining a number of the technologies outlined in this paper has the potential to enable malicious users to securely transfer any desired information to another user/machine without arousing suspicions of system administrators. Identifying the use of a HTTPS, browser-based P2P file transfer with relatively small transfer sizes might prove prohibitively difficult. The investigation of these transfers may prove cost prohibitive in terms of both time and money for law enforcement to comprehensively investigate.

\subsection{Future Work}
Pramble

\begin{itemize}
  \item Automated Detection of HTML5 based Data Exfiltration -- Unencrypted Transfers, Encrypted Transfers
  \item Approximate Hashing -- Approximate hashing facilitates the... When the blacklist gets too large and performance issues are encountered
  \item Tor Browser Plugin -- 
\end{itemize}

\bibliographystyle{plainnat}
\begin{flushleft}
\bibliography{adfsl}
\end{flushleft}

\end{document}